\documentclass[%
 reprint,
superscriptaddress,
 amsmath,amssymb,
 aps,
 prl,
 floatfix,
]{revtex4-2}

\usepackage{graphicx}
\usepackage{dcolumn}
\usepackage{bm}
\usepackage{color}
\usepackage{textcomp}
\usepackage{algpseudocodex}

\begin{document}

\preprint{APS/123-QED}

\title{Stochastically Structured Illumination Microscopy scan less super resolution imaging}

\author{Denzel Fusco} 
\affiliation{Department of Physics, University Sapienza, I-00185 Roma, Italy}
\affiliation{Center for Life Nano- \& Neuro-Science, Italian Institute of Technology, Rome, Italy}

\author{Emmanouil Xypakis}
\affiliation{Institute of Nanotechnology of the National Research Council of Italy, CNR-NANOTEC, Rome Unit, Piazzale A. Moro 5, I-00185, Rome, Italy}
\affiliation{Center for Life Nano- \& Neuro-Science, Italian Institute of Technology, Rome, Italy}

\author{Ylenia Gigante}
\affiliation{Center for Life Nano- \& Neuro-Science, Italian Institute of Technology, Rome, Italy}
\affiliation{D-Tails s.r.l. BCorp, Via di Torre Rossa, 66, 00165 Rome, Italy.}

\author{Lorenza Mautone}
\affiliation{Center for Life Nano- \& Neuro-Science, Italian Institute of Technology, Rome, Italy}
\affiliation{D-Tails s.r.l. BCorp, Via di Torre Rossa, 66, 00165 Rome, Italy.}

\author{Silvia Di Angelantonio}
\affiliation{Center for Life Nano- \& Neuro-Science, Italian Institute of Technology, Rome, Italy}
\affiliation{Department of Physiology and Pharmacology "V. Erspamer", Sapienza University of Rome, Rome, Italy.}
\affiliation{D-Tails s.r.l. BCorp, Via di Torre Rossa, 66, 00165 Rome, Italy.}

\author{Giorgia Ponsi}
\affiliation{Center for Life Nano- \& Neuro-Science, Italian Institute of Technology, Rome, Italy}
\affiliation{Department of Psychology, Sapienza University of Rome, 00185 Rome, Italy}

\author{Giancarlo Ruocco}%
\affiliation{Center for Life Nano- \& Neuro-Science, Italian Institute of Technology, Rome, Italy}
\affiliation{Department of Physics, University Sapienza, I-00185 Roma, Italy}

\author{Marco Leonetti*}
\affiliation{Institute of Nanotechnology of the National Research Council of Italy, CNR-NANOTEC, Rome Unit, Piazzale A. Moro 5, I-00185, Rome, Italy}
\affiliation{Center for Life Nano- \& Neuro-Science, Italian Institute of Technology, Rome, Italy}
\affiliation{D-Tails s.r.l. BCorp, Via di Torre Rossa, 66, 00165 Rome, Italy.}

\date{\today}

\collaboration{e-mail:  marco.leonetti@cnr.it.  Denzel Fusco and Emmanouil Xypakis equally contributed to  results presented here. }
\begin{abstract}

In Super-resolution, a varying-illumination image stack is required. This enriched the dataset typically necessitates precise mechanical control and micron scale optical alignment and repeatability. Here, we introduce a novel methodology for super-resolution microscopy called Stochastically Structured Illumination Microscopy (S$^2$IM), which bypasses the need for illumination control instead exploiting the random, uncontrolled movement of the target object. We tested our methodology within the clinically relevant ophthalmoscopic setting, harnessing the inherent saccadic motion of the eye to induce stochastic displacement of the illumination pattern on the retina. We opted to avoid human subjects by utilizing a phantom eye model, featuring a retina composed of human induced pluripotent stem cells (iPSC) retinal neurons, and replicating the ocular saccadic movements by custom actuators. Our findings demonstrate that S$^2$IM unlocks scan-less super-resolution with a resolution enhancement of 1.91, with promising prospects also beyond ophthalmoscopy applications such as active matter or atmospheric/astronomical observation.

\end{abstract}

\maketitle

\section{Introduction}

The diffraction limit, formulated almost 150 years ago by Ernst Abbe \cite{abbe1878optischen,rayleigh1896xv}, sets the maximal resolution which can be achieved with a specific lens and wavelength . This barrier can been surpassed thanks to super resolution techniques: a set of technologies merging prior knowledge on the experiment, together with multiple image acquisitions (such as sample scanning, or photoactivation cycles) to generate a single higher resolution image\cite{jing2021super}. Single molecule localization based techniques can deliver an optical resolution up to 10 nanometers\cite{betzig2006imaging,huang2008three, yamanaka2014introduction}, requiring however chemically engineered fluorochromes and activation/bleaching cycles performed with high power laser\cite{hell1994breaking}. On the other hand Structured Illumination Microscopy (SIM) based techniques, do not require the use of nonlinear optical effects, resorting instead on the additional information obtained administering to the sample a set of spatially tailored illumination \cite{chen2023superresolution, gustafsson2000surpassing},  which can now benefit of fast and artifact free algorithms \cite{10.1117/1.AP.4.2.026003,WANG2023100425} . By merging the information obtained from the multiple ($N$) measurements, it is possible to expand the support space in the frequency domain, so that the super resolved image cutoff frequency $k_{SR}$ originally equal to the collection cutoff  $k_{C}$ frequency becomes $k_{SR} \sim k_{EX} + k_{C}$, where $k_{EX}$ is the spatial cutoff frequency of the illumination pattern \cite{heintzmann2017super}.  Being the Resolution enhancement RE inversely proportional to $k_{SR}$, one argues that both by reducing the speckle grain size ($k_{EX}$), or improving the objective numerical aperture ($k_{C}$), can enhance imaging resolution \cite{strohl2016frontiers}. 
SIM is limited by the strict requirement of a fully controlled illumination pattern: the presence of aberrations or deformations in spatial structure of the illumination results in an immediate decrease of the image enhancement performance. Indeed SIM requires an external ``scanning systems'', capable to cycle between multiple illumination configurations. Several approaches have been developed to control illumination including multiple ray interferometry \cite{shao2008i5s}, microlens array \cite{orth2013gigapixel}, digital micromirror devices, \cite{brown2021multicolor} or with phase reflective masks \cite{wen2021transmission}. It is important to note that all these approaches require micro-metric alignment, and need to maintain this precision while operating at high speed as the cycling speed affects the super-resolution framerate. These illumination cycling devices, are typically expensive, light wasting, and need manutention and realignment by an expert operator.

The cost for this increased experimental complexity is typically compensated  with the possiblity to extract information from the nano-world, such as capturing images of sub-cellular structures in the range of a few tens of nanometers. This enhanced optical capability requires typcally the use of SIM togheter with the more performing optical system available i.e. immersion  objectives with the highest commercially available numerical aperture. These systems, however, operate at very close range to the sample (few hundreds of microns)  thus maximal super resolution performance cannot be reached in intrinsically long working distance experiments such as astronomic or atmospheric imaging, extreme conditions experiments requiring special chambers, (e.g., high-pressure or high-temperature chambers), or in non-invasive ophthalmoscopic environments, where the eye optical system (the cornea+crystalline systems) is employed as a $\sim$ 22 mm focal lens \cite{schwiegerling2004field,dai2008wavefront}. In these scenarios, the distance between the first optical element and the sample cannot be further reduced, resulting in images with a resolution much larger than the illumination wavelength. Thus  Structured Illumination Techniques  can improve resolution without the need of nonlinear high power pulses ( potentially harmful in live imaging or clinical contexts)  or engineered (and potentially toxic) fluorochromes.

The eye is more than just the organ responsible for vision; it serves as a direct portal to the central nervous system and is increasingly recognized for its potential in the early diagnosis of neurodegenerative diseases  \cite{mirzaei2020alzheimer, romaus2022alzheimer}.  Recent advancements have introduced specific fluorescent markers for Alzheimer’s-related protein aggregates  \cite{Boffi2022Fluorescent},  suggesting that clinical fluorescence retinal microscopy could play a significant future role in the prevention and study of neurodegenerative conditions.  Thus, the detection of these protein aggregates and subtle morphological changes in the vasculature associated with Alzheimer's and other neurodegenerative disorders necessitates the development and use of superresolved imaging techniques to enhance the resolution and clarity of retinal images \cite{grimaldi2019neuroinflammatory, gupta2021retinal, pediconi2023retinal,  nguyen2021seeing}.

Here we present a novel super resolution technique the Stochastically Structured Illumination Microscopy ($S^2IM$): a scan-less version of the structured illumination technique in which we leverage the natural and stochastic movement of the sample  to realize resolution enhancement. Our idea is an evolution of the Computational Structured Illumination Microscopy (C-SIM) which is a ``translation version'' of the  SIM. Indeed the classical SIM reported in the seminal paper from Gustafsson \cite{gustafsson2000surpassing} relies on illumination with a controlled spatial structure consisting of a periodic two dimensional parallel lines patterns in the sample plane (coordinates $[x, y]=\mathbf{r}$).  Fluorescence pattern to be reconstructed ($\rho(\mathbf{r})$) and illumination ($I(\mathbf{r})$), superpose multiplicatively, giving rise to beat patterns known as moirè fringes. Merging the information from the moirè fringes with the known shape of the illumination pattern, it is possible to reconstruct with high fidelity the spatial distribution of the fluorescent $\rho(\mathbf{r})$, which typically contains the biologically relevant information. 
Each line pattern provides information increases the Fourier support along the axis of the fluctuation, thus to reconstruct a fully super resolved image, multiple patterns with different tilt have to be delivered to the sample, so as to cover all the directions/phases in the Fourier space thus typically requiring 9 independent acquisitions \cite{gustafsson2000surpassing,perez2016optimal}.
Blind SIM revolutionized the field of linear super resolution, by providing the possibility to employ uncontrolled laser speckle patterns \cite{mudry2012structured}, thus relaxing the constraint on the illumination control. However also blind sim works under restrictive constraints such as homogeneity of the speckle grains, appearance probability, sparsity of the grains, and fully developed speckles. In a environment non-ideal for microscopy such as the human eye, it is difficult to meet these requirements:  aberrations, back reflections and retinal curvature play a role, providing a resolution enhancement much smaller than two \cite{xypakis2022deep,labouesse2017joint}. Indeed C-SIM \cite{yeh2019computational,yeh2019speckle} lies between SIM and blind SIM: it does not require a strict control on the illumination structure, but the capability to carefully translating the input pattern without inducing modification and with a controlled (and exactly known) translation amplitude. On the other hand C-SIM enables to reach the maximally achievable resolution enhancement  $RE=k_{SR}/k_{C}$ which is $\sim 2$ in the case of a single optical element employed both for illumination and collection (see \cite{yeh2019computational} for the full theoretical background of this ``translated speckles'' approach ).

$S^2IM$ is instead applicable on an ``active'' sample , i.e. an object which is characterized by an intrinsic movement. For samples characterized by intrinsic movement such as spinning astronomical elements, active matter, imaging should be  accompanied by image registration that is the process that overlays two or more images taken at different times, or from the same scene that geometrically rotates or translates. Traditional super resolution for  moving object, would be an exceptionally difficult task because one should de-multiplex images variation due to illumination scanning from variations originating from sample movement. With $S^2IM$ we turn the terms around, employing the intrinsic object movement to realize an effective scanning, and at the same time removing the need for complex and delicate scanning optical elements. Here we demonstrate the super resolution with $S^2IM$ in one of the most relevant system characterized by intrinsic movement: the human eye.

\begin{figure*}[th!]
\includegraphics[width=\textwidth]{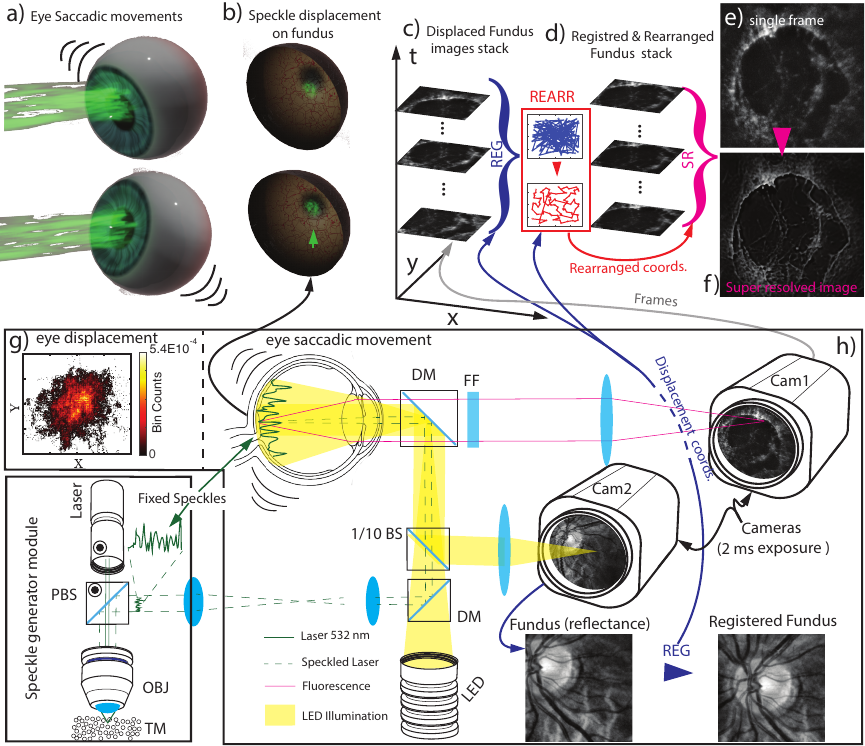}
\caption{\textbf{$\mathbf{S^2IM}$ workflow visualized and sketch of the experiment}. a-d) represent visually the various step of the $S^2IM$ data acquisition and analysisi. e) reports a typical single frame and f) the super-resolved image obtained with our super resolution approach.g) represent the probability cloud obtained with with eye measurement and replicated thanks to the m-BIME ( X and Y axis span $100 \mu m$). The  h) reports a simplified version of the experiment scheme. } 
\label{S2IM_Sketch}
\end{figure*}

\section{Results}
\subsection{Stochastic Structured Illumination Microscopy}
Fig. \ref{S2IM_Sketch} panels (a-f) reports a scheme with the basic elements of the $S^2IM$ functioning, while panel g) reports a spectch of the optical experimental setup.  Laser light  is turned into a propagating speckle pattern thought a speckle generator module ( speckle light is identified with dashed green lines in panel h),  and is delivered to the eye. The eye intrinsic saccadic and micro-saccadic movement (panel a), produce a movement of the retina with respect to the steady illumination pattern (panel b). A separate optical line provides a flat Led Illumination to the fundus (yellow illumination in panel h). The reflected pattern from the led illumination produces a reflectance image on  the camera Cam2. Being realized with a flat speckle-less illumination reflectance image can be employed for image registration. At the same time camera Cam1 retrieves images for the fluorescent signal of the eye fundus. Both camera and sources are synchronized employing a DAQ board for triggering, while exposure time is set to 2 ms to avoid motion blur induced images deformation\cite{liang1997supernormal}. The system thus produce a two stacks of images, the reflectance images stack $R(\mathbf{r},t)$ obtained with the homogeneous led illuminations and the laser generated fluorescence images stack $F(\mathbf{r},t)$.  The first set of images is employed for the registration. Being the registration process precision essentially limited by noise \cite{robinson2004fundamental,clement2018image}, by employing acquisitions counting a large number of photons (our typical reflectance imaging configuration results in  5000 photons per pixel average countrate, for images  composed of 384 $\times$ 384 pixels ), our registration process results in a sub pixels ($< 0.125$ pix  ) registration accuracy. In particular we employed freely available and optimized software dedicated to registration of rigidly translated images based on an upsampled cross correlation between images \cite{guizar2008efficient}. After  the images displacement are retrieved ($\mathbf{\Delta}(t)$), each image $F(\mathbf{r},t)$ (panel c) is translated by $-\mathbf{\Delta}(t)$, thus retrieving images  $TF(\mathbf{r},t)=\mathbf{T}(F(\mathbf{r}),-\mathbf{\Delta}(t))$ (panel d) resulting from  steady superimposed fluorescent patterns $\rho(\mathbf{r})$ and translated illuminations $I(\mathbf{r},t)=\mathbf{T}(I(\mathbf{r}),-\mathbf{\Delta}(t))$, with $\mathbf{T}$ indicating the translation operator. We thus obtain a dataset realized from a steady fluorescent object with translated illuminations

\begin{eqnarray}
\label{eq:1}
TF(\mathbf{r},t)=\mathcal{H}\otimes(\rho \times \mathbf{T}(I(\mathbf{r}),-\mathbf{\Delta}(t)))  
\end{eqnarray} 
where the $\otimes$ operator indicates convolution, and $\mathcal{H}$ is the point spread function of the collection optics.
Before applying the super resolution algorithm a further post processing step is required. Indeed for a gradient descent algorithm, it is important that the differences between frame to frame are minimal, thus only displacements smaller than the point spread function size are acceptable. On the other hand (as will be described below) the eye movement are uncontrolled and larger displacements are possible. To avoid these unwanted ``big jumps'' we rearrange the dataset temporally so that the each frame shows a small distance with respect to the previous one (red framed panel between c and d). 
After this frames rearrangement, data are passed to the $S^2IM$ gradient descent algorithm to generate the super resolved frame (panel f) from the stack of low resolution frames (an element of which is reported in panel e).
First we tested $S^2IM$ on a numerical experiment, then in  lab operative condition, simulating experimentally the ocular environment with the motorized version of the Biological Model EYE  (BIME) \cite{ferraro2023model} the m-BIME. This system is, at our knowledge the best experimental reproduction of the optical properties of the human eye which can be employed without involving patients.  The m-BIME \textit{i)} has the same optical properties of the human eye, including non-midriatic numerical aperture; \textit{ii)} is a watertight devices, thus being water-filled, also perfectly emulates the humor vitreo and humor acqueo; \textit{iii)} provides the possibility to host interchangeable biological samples, such as fixed and stained induced pluripotent stem cells (iPSC); \textit{iv)} its positioning (tilt) is controlled by two actuator dc motors with the capability to mimic human natural saccadic movement.
This last feature requires further explanation. Human patients, even during ``fixation'' experiments (in which they are asked to stare continuously an artificial target), are subject to uncontrolled, natural ocular movements which, named saccades, and which are responsible for range finding and visual acuity enhancement \cite{intoy2020finely}. Thus every patient or volunteer may have its fixation characterized with shift probability cloud, which contains information about the occupation probability of every position in the vision plane. To each position corresponds a relative retinal displacement. Indeed we measured the eye displacement, with millisecond resolution \textbf{(see methods)} on volunteers, to extract the relative retinal displacements probability cloud reported in Fig.\ref{S2IM_Sketch} panel g). The m-BIME includes indeed a software which, given a measured displacement probability cloud, replicates the eye movement by controlling the tilt motors.
Thus we can employ the m-BIME not only to simulate ocular environment but also to emulated eye movement during fixation. We employed this feature of the m-BIME to realize the $S^2IM$.

\subsection{A numerical experiment}
\begin{figure}
\includegraphics[width=8 cm]{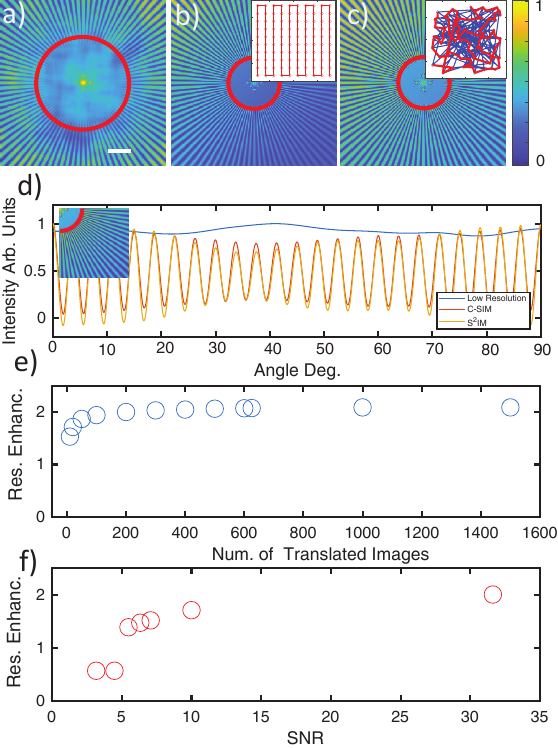}
\caption{\textbf{Numerical test of $\mathbf{S^2IM}$} Panel a) b) and c) report a Siemens star retrieved  respectively with:  diffraction limited optics, C-SIM and $S^2IM$. Insets report the qualitative arrangement of the translations position. Scale bar in A is 10 times the diffraction limited optical resolution. Inset in c) reports lines connecting points in temporal order before (blue) and after (red) the temporal rearrangement which minimizes the distance between successive positions. Panel d) reports the  Intensity versus angle (intensity along the circular profile depicted in the inset ) For Low resolution C-SIM and  $S^2IM$. Panel e) reports the resolution enhancement (Res. Enhanc.) versus the number of illumination translations (Num. of Translations), while panel d) reports the same observable versus the signal to noise ratio (SNR).} 
\label{Simulation}
\end{figure}

We first demonstrate $S^2IM$ in a numerical experiment providing with free access and control to all experimental parameters and enabling us to characterize the performance and compare it to C-SIM. We tested the two algorithms on  the Siemens Star  a sample traditionally employed to assess resolution and performances of imaging techniques (a low resolution  image of the Siemens star is reported on \ref{Simulation}a). In the numerical experiment we generated two different dataset for C-SIM and for $S^2IM$. In the C-SIM the illumination are translated on a fully deterministic and controlled pattern corresponding to a square lattice (see inset of \ref{Simulation}b) while in the $S^2IM$, illuminations are translated of a stochastic amount (see inset of \ref{Simulation}c)  (see methods for Simulations Details).    Panel \ref{Simulation}d) reports the intensity profiles along a circumference at a fixed radius for C-SIM  (red curve) and  $S^2IM$ (orange curve)  demonstrating the intrinsically similar performance. Moreover, given the symmetry of the Siemens star, it is particularly easy to determine the Resolution Enhancement (see \cite{xypakis2022deep} and supplementary information's). In panel \ref{Simulation}e, we reported the RE versus the number of acquired images, demonstrating how the technique start a saturation  to the value of 2 when more than 200 frames are acquired. Panel \ref{Simulation}f instead reports about the robustness of $S^2IM$ with respect to noise. In the numerical experiment synthetic noise (Poissonian shot noise and readout noise) is added to the data in order to assess the RE. In the supplementary materials, we also report on the robustness of the technique with respect to the error on the position retrieval, demonstrating that our registration error ($< 0.125$ pix ) is much smaller than the value at which $S^2IM$ performance results to be affected (1 pix ).

\section{Results on a test target}
We tested our technique on a test target made of 15 $\mu m$ diameter, shell-stained beads (Invitrogen Focalcheck F7235)  randomly dispersed on a 12 mm circular cover-slip mounted on the m-BIME.  In clinical experiments, the camera exposure cannot exceed the 4 ms in order to avoid, eye-movement induced motion blur \cite{liang1997supernormal}. In a conservative framework we set the exposure time of all the experiments (both test target and retinal cultures) to the even more restrictive value of 2 ms.

These shell-stained beads produce a donut-like signal, which results into full disks due to the limited resolution of the ocular environment (system resolution is set by the eye model to 6.5 $\mu m$\cite{bakaraju2008finite}).

\begin{figure}
\includegraphics[width=8 cm]{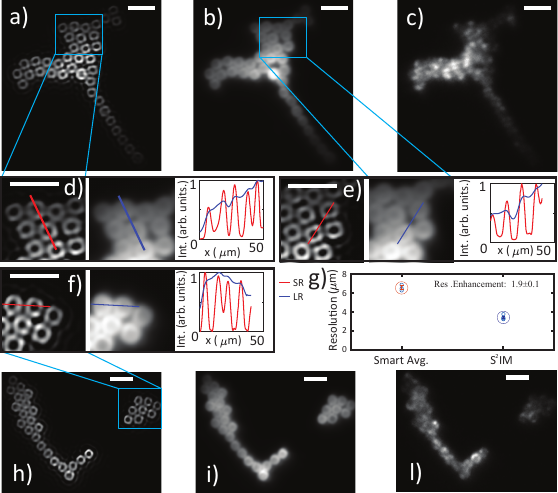}
\caption{\textbf{$\mathbf{S^2IM}$ on a test sample.} Two fields containing 15 $\mu m$ diameter fluorescent donuts. Panels a) b) and c)  represent  $S^2IM$, smart averaged (realigned and than averaged), and single shot images of the same filed. The same is for panels h) i) and l) for a different field. 
Panels d-f) represent zoom in of the areas indicated by blue rectangles. Each of these panels is organized in three sub panels reporting  $S^2IM$, smart averaging  and intensity profiles along the blue-red lines. Panel g) reports Resolution extracted from the images. Scale bars are 40 $\mu m$.  } 
\label{Donuts}
\end{figure}

Fig. \ref{Donuts} report two fields. In panels a, b, c we report respectively, the super resolved, the realigned and averaged (smart averaging ) and the single shot images for the first field. \ref{Donuts}a and \ref{Donuts}b result from 2000 acquisitions each with a 2ms exposure time. In \ref{Donuts}c intensity fluctuations due to speckled illumination are distinguishable. Panels \ref{Donuts}d-e) contains comparison between $S^2IM$ and smart-averaged images for two zoom in areas and relative intensity profiles along the highlighted lines. Panels \ref{Donuts}h-l)  and \ref{Donuts}f) are organized in the same way as the previous but represent a different field. Panel \ref{Donuts}g)report the retrieved resolution estimate for smart-averaged and super resolved images.

 The estimation of resolution in the two cases is assessed as the smallest closer distinguishable (Rayleigh criterion, i.e. peaks separated by at least a 73.5 \% less intense dip \cite{SHEPPARD200561}) donuts boundaries distance. With this approach we retrieve a 6.5 $\pm 0.2  \mu m$ resolution for the wide field images, and a  3.4 $\pm 0.1  \mu m$ for the  $S^2IM$ super resolved images. Average and errors are obtained from an estimate on 10 different imaging fields. The resolution enhancement is thus RE $=  1.9  \pm 0.1 $, in agreement with the estimate from numerical simulations from the Siemens star.


\subsection{Results on a retinal neurons culture}
\begin{figure*}
\includegraphics[width=\textwidth]{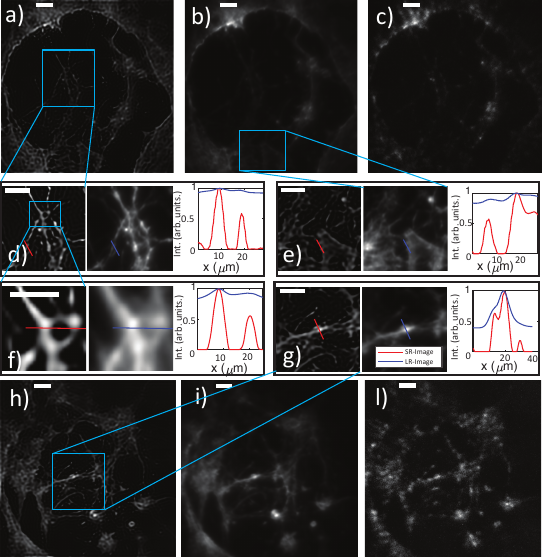}
\caption{ \textbf{$\mathbf{S^2IM}$ on retinal neurons.} Images relative to retinal neurons culture actine-stained and loaded on the m-BIME.  Panels a-c) Report (from left to right), $S^2IM$, smart averaging, and single shot on the first field. Panels a-c) Report (from left to right), $S^2IM$, smart averaging, and single shot. Panels h-l) are organized as panels a-c). Panels d-g) represent zoom in of the areas highlighted by the blue squares.  Each of these panels is organized in three sub panels reporting  $S^2IM$, smart averaging and intensity profiles along the blue-red lines. Exposure time is 2ms, and $N = 2000$ single shot images have been acquired. Scale bars are all by 40 $\mu m$ except for panel f) which is 20 $\mu m$. } 
\label{Neurons}
\end{figure*}

We tested $S^2IM$  human iPSC-derived retinal neurons (see methods) stained with Phalloidin-Atto 532. Note that the signal from this this biological sample is from 10 to 3 times less intense than the one produced by synthetic fluorescent donuts of our test target. This summed to the reduced exposure time needed to avoid motion blur in the eye (2 milliseconds, \cite{liang1997supernormal}, 50-100 times smaller than the typical exposure time employed for fluorescence microscopy experiments) and to the reduced luminosity of the collection optics (numerical aperture of $\sim$ 0.04), makes this imaging task extremely challenging, thus furhter supporting the importance of the presented results. Fig. \ref{Neurons} reports two different field of view comparing $S^2IM$, smart averaging and single shot images, highlighting the level of details which can be obtained thanks to our technique. A larger set of images is reported in \textbf{supplementary materials}.

\section{Discussion}
$S^2IM$ not only provides platform free from sophisticated and costly scanning hardware, but also provides an avenue to super resolve moving objects which are typically inaccessible because of the intrinsic complexity of the multiplexing illumination scanning from uncontrolled movement. 
This scan-less techniques can reach the same theoretical resolution enhancement of standard SIM experiments even if the uncontrolled sample is characterized by unpredictable movements. We demonstrated $S^2IM$ in the clinically relevant environment of the human eye, but it can be potentially exported to other important fields such as the distance imaging (atmospheric, aeronautics or astronomic) or to other biological system characterized by unpredictable intrinsic motion such as active matter system.

\section{Methods}

\textit{Ethics and Informed consent}  All procedures performed in studies involving human participants were in accordance with the ethical standards of the institutional and/or national research committee and with the 1964 Helsinki Declaration and its later amendments or comparable ethical standards. The study was approved by the ``Independent ethical committee'' of the Fondazione Santa Lucia IRCCS Via Ardeatina 306, 00179 Roma, Italy  (Prot.  CE/2023$\_$023  of the 04/04/23).  Informed consent was obtained from all human research participants. \textit{iPSCs use:}  We used commercially available human iPSCs provided by the EBISC biobank (https://ebisc.org/) (CERT approval 5/2022 to SDA; https://www.uniroma1.it/en/pagina/ethics-committee-transdisciplinary-research). Informed consent was not needed for this aspect of the research.
\\
\textit{Data availability} Experimental and generated data related to the generated in this study  are deposited in the GitHub repository \href{https://github.com/emmxyp/super-resolution-algorithm}{$https://github.com/emmxyp/super-resolution-algorithm$} .
\\
\textit{Code availability} Code realized in this study  are deposited in the GitHub repository at the address  \href{https://github.com/emmxyp/super-resolution-algorithm}{$https://github.com/emmxyp/super-resolution-algorithm$}.
\\
\textit{Aknowledgements:} This work was supported by MUR PRIN 2022 (CUP:  2022CFP7RF, to SDA and ML). This research was also funded by the D-Tails-IIT Joint Lab (to SDA, YG, ML), the Regione Lazio FSE 2014–2020 (19036AP000000019 and A0112E0073) grants (to SDA), Sapienza University grants (RM118163E0297F84, PH12017270934C3C, and MA32117A7B698029 to SDA), and Fondazione Istituto Italiano di Tecnologia (to LM). LM was also supported by the PhD program in Life Science at Sapienza University in Rome. SDA was also supported by Progetto ECS 0000024 Rome Technopole, - CUP B83C22002820006, PNRR Missione 4 Componente 2 Investimento 1.5, finanziato dall’Unione europea – NextGenerationEU. GR iso supported by  Project ``National Center for Gene Therapy and Drugs based on RNA Technology'' (CN00000041) financed by NextGenerationEU PNRR MUR—M4C2—Action 1.4-Call ``Potenziamento strutture di ricerca e creazione di ``campioni nazionali di R\&S" (CUP J33C22001130001)''. The research leading to these results was also supported by European Research Council through its Synergy grant program, project ASTRA (grant agreement No 855923) and by European Innovation Council through its Pathfinder Open Programme, project ivBM-4PAP (grant agreement No 101098989). 
\\
\textit{Author contributions:}  M.L. and G.R. Conceived the Idea and designed the S$^2$IM device,  M.L. and  D.F. Realized the device. D.F. Performed the experiments and realized the preliminary data analysis. E.X. realized and optimized the code for the Images enhancement and super resolution and performed data analysis and resolution estimate simulations. G.P. realized the Saccadic movement measurements. S.D.A., L.M. and Y.G. realized the Human iPSCs cell  Cultures, Fixation and staining. 
\\
\textit{Competing Interests:}  The authors declare no competing interests.

\subsection{Mesurement of the retinal saccadic movements}
Volunteers took part in a fixation task, while their binocular eye movements were recorded using a high-resolution (spatial accuracy: 0.25-0.50$^\circ$; resolution: 0.01$^\circ$) infrared eye tracker (sampling rate: up to 2000 Hz;  EyeLink \textregistered  1000 Plus, SR Research). Head movements were restrained. Built-in calibration and validation procedures were employed to ensure acceptable spatial accuracy (maximum error for each point $<$ $1^\circ$; average error $<$ $0.5^\circ$). 

Each trial started with a central black fixation cross (0.53$^\circ$ x 0.28$^\circ$) that participants had to look for at least 300 ms, to initiate the task. Afterwards, a black fixation dot (0.27$^\circ$ x 0.27$^\circ$) appeared at the centre of the screen for a variable time. Participants were instructed to look at the dot until its disappearance, by maintaining fixation as much as possible. Finally, a blank screen was presented for 100 ms. The order of trials (n = 15) was randomized. The methodology followed the latest reporting guidelines in eye tracking research (\cite{dunn2023minimal}).

\subsection{Human iPSCs differentiation into retinal neuron} 
Retinal cultures were differentiated from healthy human induced-pluripotent stem cell line (iPSC) according to a previously published protocol with minor modifications \cite{sluch2017enhanced}. Human iPSCs were dissociated to single cells with Accutase (Gibco), and seeded into Matrigel-coated dishes (Corning, dilution 1:100) at a density of 100.000 cells per cm 2 . The day of seeding the cells were maintained in mTeSR Plus with 10 µM Rock Inhibitor (RI; Peprotech, Cranbury, New Jersey, Stati Uniti). One day after seeding, the medium is switched to neurogenic basal medium (N2B27w/oA) composed of of 50\% DMEM/F12 (Sigma), 50\% Neurobasal (ThermoFisher), 1\% GlutaMAX Supplement (Gibco), 0.1\% Pen-Strep (Sigma), 1\% NEAA (Gibco), 1\% N2 Supplement (ThermoFisher), and 2\% B27 Supplement w/oA (Gibco). A different mix of small molecules was added to the medium at specific intervals to induce retinal progenitor induction (DIV 20, days in vitro) and differentiation into retinal neurons up to 30 days: 1 µM Dorsomorphin (Sigma) and 2.5 µM IDE2 (Sigma) from DIV 0 to DIV 6; 10 mM Nicotinamide (Sigma) until DIV 10; 25 µM Forskolin (Sigma) from DIV 0 to DIV 30; 10 µM DAPT (Prepotech) from DIV 20 to DIV 30. The retinal progenitor cells were plated onto PLO/Laminin (Sigma-Aldrich) coated round cover glasses (Ø 12mm, Thorlabs, Newton, New Jersey, US) at a density of 150.000 cells per glass for subsequent analysis.

\subsection{Immunostaining} 
iPSC-derived retinal neurons were fixed at DIV 30 of differentiation with 4\% paraformaldehyde (PFA, Sigma Aldrich) for 15 minutes at room temperature. Fixed retinal cells were permeabilized using PBS solution with 0.2\% Triton X-100 (Sigma Aldrich) for 10 minutes and incubated in blocking solution containing PBS, 0.1\% Tween-20 (Sigma-Aldrich) and 5\% goat serum (Merck KGaA, Darmstadt, Germany) for 45 minutes. Subsequently, retinal neurons were incubated with Phalloidin-Atto 532 (1:30, 49429. Sigma Aldrich) for 1h at room temperature. Hoechst was used to stain nuclei (1:300, 33258, Merck). The retinal cells were sealed between two cover glasses using DAKO (Fluorescent Mounting Medium, Sigma-Aldrich) and mounted on the m-BIME.

\subsection{Super-resolution algorithm}
After the steady fluorescent object dataset is obtained as described in the main text (Eq. \ref{eq:1}) the probrem is reduced to the following.
A flourophore density $\rho(\mathbf{r})$ is illuminated by an unknown intensity profile  $I_\Lambda(\mathbf{r}) $ by a set of $\Lambda= 1,2,3 ...N$ consequent measurements. Each measurement differs by the previous by the fact that the illumination pattern on the sample plane $r$ is translated by a distance $\mathbf{r}_{\Lambda}$ from the first measurement so that $I_\Lambda(\mathbf{r}) =I(\mathbf{r + r_\Lambda}) $, where $I(\mathbf{r})$  is the illumination pattern of the first measurement. The image is formatted on the camera image plane $\mathbf{y}$ and captured by a detector $D(\mathbf{y})$ so that for each measurement $\Lambda$.

\begin{equation}
    D_\Lambda(\mathbf{y}) = \sum_{\mathbf{x}} h(\mathbf{x}, \mathbf{y}) \rho(\mathbf{x}) I_\Lambda (\mathbf{x}) \
\end{equation}
where $h(\mathbf{x}, \mathbf{y})$ is the  point spread function of the optical setup.

The task of the super-resolution algorithm is to find an approximation $\tilde{\rho}(\mathbf{x})$ and $\tilde{I}(\mathbf{x})$ by minimizing a loss function 
\begin{equation}
    L( \mathbf{x})= \sum_\mathbf{y} \mathcal{L}_\Lambda ( D_\Lambda(\mathbf{y}) ,  \tilde{D}_\Lambda(\mathbf{y})  )
\end{equation}
in an iterating process assuming that the low resolution image that the algorithm produces is .

The fluorophore density guess $\tilde{\rho}(\mathbf{x})$ and the intensity guess $\tilde{I}_\Lambda(\mathbf{x})$ are updated iterative for a number of steps $s$

\begin{equation}
    \tilde{\rho}_{s+1}(\mathbf{x}) =  \tilde{\rho}_{s}(\mathbf{x}) - \alpha \mathbf{\partial}_{\tilde{\rho}_{s}} L 
\end{equation}

and 

\begin{equation}
    \tilde{I}_{s+1}(\mathbf{x}) =  \tilde{I}_{s}(\mathbf{x}) - \alpha \mathbf{\partial}_{\tilde{I}_{s}} L 
\end{equation}

where $\alpha$ is the gradient descent step where we take to be $\alpha = 0.01$ and 
\begin{equation}
    \mathbf{\partial}_{\tilde{\rho}_{s}} L = \sum_\mathbf{y} (\mathbf{\partial}_{\tilde{D}_\Lambda} \mathcal{L}_\Lambda ) (\mathbf{\partial}_{\tilde{\rho}_{s}}{\tilde{D}_\Lambda})
\end{equation}

\begin{equation}
    \mathbf{\partial}_{\tilde{I}_{s}} L = 
    \sum_\mathbf{y} (\mathbf{\partial}_{\tilde{D}_\Lambda} \mathcal{L}_\Lambda ) (\mathbf{\partial}_{\tilde{I}_{s}}{\tilde{D}_\Lambda})
\end{equation}

For the low photon noise experimental data we make use of the Kullback Leibler divergence \cite{Xypakis:23}

\begin{equation}
    \mathcal{L}_\Lambda =  \tilde{D}_\Lambda \log{\frac{\tilde{D}_\Lambda}{D_\Lambda}} + D_\Lambda-\tilde{D}_\Lambda
\end{equation}
A simplified pseudocode for the algorithm is the following
\begin{algorithmic}
\For{s=1: S} 

\For{$\Lambda$ =1 :N} 

\State $\tilde{D}_\Lambda = \mathcal{H}\otimes (\tilde{\rho} \tilde{I}_\Lambda) $
\State $\tilde{\rho} \gets \tilde{\rho} - \alpha \tilde{I}_\Lambda \mathcal{H}\otimes  (\log{\frac{\tilde{D}}{D}}) $

\State $\tilde{I}_\Lambda \gets \tilde{I}_\Lambda - \alpha \tilde{\rho} \mathcal{H}\otimes  (\log{\frac{\tilde{D}}{D}}) $
\EndFor
\EndFor

\end{algorithmic}

An user friendly version of the algorithm can be found in \href{ https://github.com/emmxyp/super-resolution-algorithm}{https://github.com/emmxyp/super-resolution-algorithm}

For the numerical simulations we use the experimental relevant parameters:
numerical aperture $NA = 0.04$, wavelength $\lambda = 0.605 \mu m$, magnification $m = 4.6875$, camera pixel size  $px = 6.5 \mu m$, field of view $FOV = 531.1061 \mu m$. Both for the C-SIM and S2IM we use 625 translations. The translation step for the C-SIM was taken to be $2.7734 \mu m$. 

The processing time for a 384x384x2000 acquisition stack is about 6 minutes (25 iterations), on a CPU AMD Ryzen 9 7900X3D and Nvidia GTX 1080 GPU. The code has been developed in Matlab, in a non parallel format. Rough estimate indicate that processing time can be reduced a factor 10/50 with parallelization and optimization. Quasi-live imaging (1-5 s processing time), can be eventually be realized with an optimized code for limited size stacks (384x384x200).

\bibliography{apssamp}

\providecommand{\noopsort}[1]{}\providecommand{\singleletter}[1]{#1}%
\begin{thebibliography}{10}
\expandafter\ifx\csname url\endcsname\relax
  \def\url#1{\texttt{#1}}\fi
\expandafter\ifx\csname urlprefix\endcsname\relax\def\urlprefix{URL }\fi
\providecommand{\bibinfo}[2]{#2}
\providecommand{\eprint}[2][]{\url{#2}}

\bibitem{abbe1878optischen}
\bibinfo{author}{Abbe, E.}
\newblock \emph{\bibinfo{title}{Die optischen H{\"u}lfsmittel der Mikroskopie}} (\bibinfo{publisher}{Vieweg}, \bibinfo{year}{1878}).

\bibitem{rayleigh1896xv}
\bibinfo{author}{Rayleigh}.
\newblock \bibinfo{title}{Xv. on the theory of optical images, with special reference to the microscope}.
\newblock \emph{\bibinfo{journal}{The London, Edinburgh, and Dublin Philosophical Magazine and Journal of Science}} \textbf{\bibinfo{volume}{42}}, \bibinfo{pages}{167--195} (\bibinfo{year}{1896}).

\bibitem{jing2021super}
\bibinfo{author}{Jing, Y.}, \bibinfo{author}{Zhang, C.}, \bibinfo{author}{Yu, B.}, \bibinfo{author}{Lin, D.} \& \bibinfo{author}{Qu, J.}
\newblock \bibinfo{title}{Super-resolution microscopy: shedding new light on in vivo imaging}.
\newblock \emph{\bibinfo{journal}{Frontiers in Chemistry}} \textbf{\bibinfo{volume}{9}}, \bibinfo{pages}{746900} (\bibinfo{year}{2021}).

\bibitem{betzig2006imaging}
\bibinfo{author}{Betzig, E.} \emph{et~al.}
\newblock \bibinfo{title}{Imaging intracellular fluorescent proteins at nanometer resolution}.
\newblock \emph{\bibinfo{journal}{science}} \textbf{\bibinfo{volume}{313}}, \bibinfo{pages}{1642--1645} (\bibinfo{year}{2006}).

\bibitem{huang2008three}
\bibinfo{author}{Huang, B.}, \bibinfo{author}{Wang, W.}, \bibinfo{author}{Bates, M.} \& \bibinfo{author}{Zhuang, X.}
\newblock \bibinfo{title}{Three-dimensional super-resolution imaging by stochastic optical reconstruction microscopy}.
\newblock \emph{\bibinfo{journal}{Science}} \textbf{\bibinfo{volume}{319}}, \bibinfo{pages}{810--813} (\bibinfo{year}{2008}).

\bibitem{yamanaka2014introduction}
\bibinfo{author}{Yamanaka, M.}, \bibinfo{author}{Smith, N.~I.} \& \bibinfo{author}{Fujita, K.}
\newblock \bibinfo{title}{Introduction to super-resolution microscopy}.
\newblock \emph{\bibinfo{journal}{Microscopy}} \textbf{\bibinfo{volume}{63}}, \bibinfo{pages}{177--192} (\bibinfo{year}{2014}).

\bibitem{hell1994breaking}
\bibinfo{author}{Hell, S.~W.} \& \bibinfo{author}{Wichmann, J.}
\newblock \bibinfo{title}{Breaking the diffraction resolution limit by stimulated emission: stimulated-emission-depletion fluorescence microscopy}.
\newblock \emph{\bibinfo{journal}{Optics letters}} \textbf{\bibinfo{volume}{19}}, \bibinfo{pages}{780--782} (\bibinfo{year}{1994}).

\bibitem{chen2023superresolution}
\bibinfo{author}{Chen, X.} \emph{et~al.}
\newblock \bibinfo{title}{Superresolution structured illumination microscopy reconstruction algorithms: a review}.
\newblock \emph{\bibinfo{journal}{Light: Science \& Applications}} \textbf{\bibinfo{volume}{12}}, \bibinfo{pages}{172} (\bibinfo{year}{2023}).

\bibitem{gustafsson2000surpassing}
\bibinfo{author}{Gustafsson, M.~G.}
\newblock \bibinfo{title}{Surpassing the lateral resolution limit by a factor of two using structured illumination microscopy}.
\newblock \emph{\bibinfo{journal}{Journal of microscopy}} \textbf{\bibinfo{volume}{198}}, \bibinfo{pages}{82--87} (\bibinfo{year}{2000}).

\bibitem{10.1117/1.AP.4.2.026003}
\bibinfo{author}{Wang, Z.} \emph{et~al.}
\newblock \bibinfo{title}{{High-speed image reconstruction for optically sectioned, super-resolution structured illumination microscopy}}.
\newblock \emph{\bibinfo{journal}{Advanced Photonics}} \textbf{\bibinfo{volume}{4}}, \bibinfo{pages}{026003} (\bibinfo{year}{2022}).
\newblock \urlprefix\url{https://doi.org/10.1117/1.AP.4.2.026003}.

\bibitem{WANG2023100425}
\bibinfo{author}{Wang, Z.} \emph{et~al.}
\newblock \bibinfo{title}{Rapid, artifact-reduced, image reconstruction for super-resolution structured illumination microscopy}.
\newblock \emph{\bibinfo{journal}{The Innovation}} \textbf{\bibinfo{volume}{4}}, \bibinfo{pages}{100425} (\bibinfo{year}{2023}).
\newblock \urlprefix\url{https://www.sciencedirect.com/science/article/pii/S266667582300053X}.

\bibitem{heintzmann2017super}
\bibinfo{author}{Heintzmann, R.} \& \bibinfo{author}{Huser, T.}
\newblock \bibinfo{title}{Super-resolution structured illumination microscopy}.
\newblock \emph{\bibinfo{journal}{Chemical reviews}} \textbf{\bibinfo{volume}{117}}, \bibinfo{pages}{13890--13908} (\bibinfo{year}{2017}).

\bibitem{strohl2016frontiers}
\bibinfo{author}{Str{\"o}hl, F.} \& \bibinfo{author}{Kaminski, C.~F.}
\newblock \bibinfo{title}{Frontiers in structured illumination microscopy}.
\newblock \emph{\bibinfo{journal}{Optica}} \textbf{\bibinfo{volume}{3}}, \bibinfo{pages}{667--677} (\bibinfo{year}{2016}).

\bibitem{shao2008i5s}
\bibinfo{author}{Shao, L.} \emph{et~al.}
\newblock \bibinfo{title}{I5s: wide-field light microscopy with 100-nm-scale resolution in three dimensions}.
\newblock \emph{\bibinfo{journal}{Biophysical journal}} \textbf{\bibinfo{volume}{94}}, \bibinfo{pages}{4971--4983} (\bibinfo{year}{2008}).

\bibitem{orth2013gigapixel}
\bibinfo{author}{Orth, A.} \& \bibinfo{author}{Crozier, K.}
\newblock \bibinfo{title}{Gigapixel fluorescence microscopy with a water immersion microlens array}.
\newblock \emph{\bibinfo{journal}{Optics express}} \textbf{\bibinfo{volume}{21}}, \bibinfo{pages}{2361--2368} (\bibinfo{year}{2013}).

\bibitem{brown2021multicolor}
\bibinfo{author}{Brown, P.~T.}, \bibinfo{author}{Kruithoff, R.}, \bibinfo{author}{Seedorf, G.~J.} \& \bibinfo{author}{Shepherd, D.~P.}
\newblock \bibinfo{title}{Multicolor structured illumination microscopy and quantitative control of polychromatic light with a digital micromirror device}.
\newblock \emph{\bibinfo{journal}{Biomedical Optics Express}} \textbf{\bibinfo{volume}{12}}, \bibinfo{pages}{3700--3716} (\bibinfo{year}{2021}).

\bibitem{wen2021transmission}
\bibinfo{author}{Wen, K.} \emph{et~al.}
\newblock \bibinfo{title}{Transmission structured illumination microscopy for quantitative phase and scattering imaging}.
\newblock \emph{\bibinfo{journal}{Frontiers in Physics}} \textbf{\bibinfo{volume}{8}}, \bibinfo{pages}{630350} (\bibinfo{year}{2021}).

\bibitem{schwiegerling2004field}
\bibinfo{author}{Schwiegerling, J.} \emph{et~al.}
\newblock \bibinfo{title}{Field guide to visual and ophthalmic optics}.
\newblock In \emph{\bibinfo{booktitle}{Field guide to visual and ophthalmic optics}} (\bibinfo{organization}{Spie Bellingham, Washington, USA}, \bibinfo{year}{2004}).

\bibitem{dai2008wavefront}
\bibinfo{author}{Dai, G.-m.}
\newblock \emph{\bibinfo{title}{Wavefront optics for vision correction}}, vol. \bibinfo{volume}{179} (\bibinfo{publisher}{SPIE press}, \bibinfo{year}{2008}).

\bibitem{mirzaei2020alzheimer}
\bibinfo{author}{Mirzaei, N.} \emph{et~al.}
\newblock \bibinfo{title}{Alzheimer’s retinopathy: seeing disease in the eyes}.
\newblock \emph{\bibinfo{journal}{Frontiers in neuroscience}} \textbf{\bibinfo{volume}{14}}, \bibinfo{pages}{921} (\bibinfo{year}{2020}).

\bibitem{romaus2022alzheimer}
\bibinfo{author}{Romaus-Sanjurjo, D.} \emph{et~al.}
\newblock \bibinfo{title}{Alzheimer’s disease seen through the eye: ocular alterations and neurodegeneration}.
\newblock \emph{\bibinfo{journal}{International Journal of Molecular Sciences}} \textbf{\bibinfo{volume}{23}}, \bibinfo{pages}{2486} (\bibinfo{year}{2022}).

\bibitem{Boffi2022Fluorescent}
\bibinfo{author}{Boffi, A.}, \bibinfo{author}{Ghirga, F.}, \bibinfo{author}{Solopert, A.} \& \bibinfo{author}{Di~Angelantonio, S.}
\newblock \bibinfo{title}{Fluorescent markers for neurofibrillar tangles and uses thereof} (\bibinfo{year}{2022}).
\newblock \urlprefix\url{https://patents.google.com/patent/WO2022224151A1}.
\newblock \bibinfo{note}{Wo 2022/224151 A1}.

\bibitem{grimaldi2019neuroinflammatory}
\bibinfo{author}{Grimaldi, A.} \emph{et~al.}
\newblock \bibinfo{title}{Neuroinflammatory processes, a1 astrocyte activation and protein aggregation in the retina of alzheimer’s disease patients, possible biomarkers for early diagnosis}.
\newblock \emph{\bibinfo{journal}{Frontiers in neuroscience}} \textbf{\bibinfo{volume}{13}}, \bibinfo{pages}{925} (\bibinfo{year}{2019}).

\bibitem{gupta2021retinal}
\bibinfo{author}{Gupta, V.~B.} \emph{et~al.}
\newblock \bibinfo{title}{Retinal changes in alzheimer's disease—integrated prospects of imaging, functional and molecular advances}.
\newblock \emph{\bibinfo{journal}{Progress in retinal and eye research}} \textbf{\bibinfo{volume}{82}}, \bibinfo{pages}{100899} (\bibinfo{year}{2021}).

\bibitem{pediconi2023retinal}
\bibinfo{author}{Pediconi, N.} \emph{et~al.}
\newblock \bibinfo{title}{Retinal fingerprints of als in patients: Ganglion cell apoptosis and tdp-43/p62 misplacement}.
\newblock \emph{\bibinfo{journal}{Frontiers in Aging Neuroscience}} \textbf{\bibinfo{volume}{15}}, \bibinfo{pages}{1110520} (\bibinfo{year}{2023}).

\bibitem{nguyen2021seeing}
\bibinfo{author}{Nguyen, C.~T.}, \bibinfo{author}{Acosta, M.~L.}, \bibinfo{author}{Di~Angelantonio, S.} \& \bibinfo{author}{Salt, T.~E.}
\newblock \bibinfo{title}{Seeing beyond the eye: the brain connection}.
\newblock \emph{\bibinfo{journal}{Frontiers in neuroscience}} \textbf{\bibinfo{volume}{15}}, \bibinfo{pages}{719717} (\bibinfo{year}{2021}).

\bibitem{perez2016optimal}
\bibinfo{author}{Perez, V.}, \bibinfo{author}{Chang, B.-J.} \& \bibinfo{author}{Stelzer, E. H.~K.}
\newblock \bibinfo{title}{Optimal 2d-sim reconstruction by two filtering steps with richardson-lucy deconvolution}.
\newblock \emph{\bibinfo{journal}{Scientific reports}} \textbf{\bibinfo{volume}{6}}, \bibinfo{pages}{37149} (\bibinfo{year}{2016}).

\bibitem{mudry2012structured}
\bibinfo{author}{Mudry, E.} \emph{et~al.}
\newblock \bibinfo{title}{Structured illumination microscopy using unknown speckle patterns}.
\newblock \emph{\bibinfo{journal}{Nature Photonics}} \textbf{\bibinfo{volume}{6}}, \bibinfo{pages}{312--315} (\bibinfo{year}{2012}).

\bibitem{xypakis2022deep}
\bibinfo{author}{Xypakis, E.} \emph{et~al.}
\newblock \bibinfo{title}{Deep learning for blind structured illumination microscopy}.
\newblock \emph{\bibinfo{journal}{Scientific Reports}} \textbf{\bibinfo{volume}{12}}, \bibinfo{pages}{8623} (\bibinfo{year}{2022}).

\bibitem{labouesse2017joint}
\bibinfo{author}{Labouesse, S.} \emph{et~al.}
\newblock \bibinfo{title}{Joint reconstruction strategy for structured illumination microscopy with unknown illuminations}.
\newblock \emph{\bibinfo{journal}{IEEE Transactions on Image Processing}} \textbf{\bibinfo{volume}{26}}, \bibinfo{pages}{2480--2493} (\bibinfo{year}{2017}).

\bibitem{yeh2019computational}
\bibinfo{author}{Yeh, L.-H.}, \bibinfo{author}{Chowdhury, S.} \& \bibinfo{author}{Waller, L.}
\newblock \bibinfo{title}{Computational structured illumination for high-content fluorescence and phase microscopy}.
\newblock \emph{\bibinfo{journal}{Biomedical optics express}} \textbf{\bibinfo{volume}{10}}, \bibinfo{pages}{1978--1998} (\bibinfo{year}{2019}).

\bibitem{yeh2019speckle}
\bibinfo{author}{Yeh, L.-H.}, \bibinfo{author}{Chowdhury, S.}, \bibinfo{author}{Repina, N.~A.} \& \bibinfo{author}{Waller, L.}
\newblock \bibinfo{title}{Speckle-structured illumination for 3d phase and fluorescence computational microscopy}.
\newblock \emph{\bibinfo{journal}{Biomedical optics express}} \textbf{\bibinfo{volume}{10}}, \bibinfo{pages}{3635--3653} (\bibinfo{year}{2019}).

\bibitem{liang1997supernormal}
\bibinfo{author}{Liang, J.}, \bibinfo{author}{Williams, D.~R.} \& \bibinfo{author}{Miller, D.~T.}
\newblock \bibinfo{title}{Supernormal vision and high-resolution retinal imaging through adaptive optics}.
\newblock \emph{\bibinfo{journal}{JOSA A}} \textbf{\bibinfo{volume}{14}}, \bibinfo{pages}{2884--2892} (\bibinfo{year}{1997}).

\bibitem{robinson2004fundamental}
\bibinfo{author}{Robinson, D.} \& \bibinfo{author}{Milanfar, P.}
\newblock \bibinfo{title}{Fundamental performance limits in image registration}.
\newblock \emph{\bibinfo{journal}{IEEE Transactions on Image Processing}} \textbf{\bibinfo{volume}{13}}, \bibinfo{pages}{1185--1199} (\bibinfo{year}{2004}).

\bibitem{clement2018image}
\bibinfo{author}{Clement, C.~B.}, \bibinfo{author}{Bierbaum, M.} \& \bibinfo{author}{Sethna, J.~P.}
\newblock \bibinfo{title}{Image registration and super resolution from first principles}.
\newblock \emph{\bibinfo{journal}{arXiv preprint arXiv:1809.05583}}  (\bibinfo{year}{2018}).

\bibitem{guizar2008efficient}
\bibinfo{author}{Guizar-Sicairos, M.}, \bibinfo{author}{Thurman, S.~T.} \& \bibinfo{author}{Fienup, J.~R.}
\newblock \bibinfo{title}{Efficient subpixel image registration algorithms}.
\newblock \emph{\bibinfo{journal}{Optics letters}} \textbf{\bibinfo{volume}{33}}, \bibinfo{pages}{156--158} (\bibinfo{year}{2008}).

\bibitem{ferraro2023model}
\bibinfo{author}{Ferraro, G.} \emph{et~al.}
\newblock \bibinfo{title}{A model eye for fluorescent characterization of retinal cultures and tissues}.
\newblock \emph{\bibinfo{journal}{Scientific Reports}} \textbf{\bibinfo{volume}{13}}, \bibinfo{pages}{10983} (\bibinfo{year}{2023}).

\bibitem{intoy2020finely}
\bibinfo{author}{Intoy, J.} \& \bibinfo{author}{Rucci, M.}
\newblock \bibinfo{title}{Finely tuned eye movements enhance visual acuity}.
\newblock \emph{\bibinfo{journal}{Nature communications}} \textbf{\bibinfo{volume}{11}}, \bibinfo{pages}{795} (\bibinfo{year}{2020}).

\bibitem{bakaraju2008finite}
\bibinfo{author}{Bakaraju, R.~C.}, \bibinfo{author}{Ehrmann, K.}, \bibinfo{author}{Papas, E.} \& \bibinfo{author}{Ho, A.}
\newblock \bibinfo{title}{Finite schematic eye models and their accuracy to in-vivo data}.
\newblock \emph{\bibinfo{journal}{Vision research}} \textbf{\bibinfo{volume}{48}}, \bibinfo{pages}{1681--1694} (\bibinfo{year}{2008}).

\bibitem{SHEPPARD200561}
\bibinfo{author}{Sheppard, C.}
\newblock \bibinfo{title}{Microscopy | overview}.
\newblock In \bibinfo{editor}{Guenther, R.~D.} (ed.) \emph{\bibinfo{booktitle}{Encyclopedia of Modern Optics}}, \bibinfo{pages}{61--69} (\bibinfo{publisher}{Elsevier}, \bibinfo{address}{Oxford}, \bibinfo{year}{2005}).
\newblock \urlprefix\url{https://www.sciencedirect.com/science/article/pii/B012369395000823X}.

\bibitem{dunn2023minimal}
\bibinfo{author}{Dunn, M.~J.} \emph{et~al.}
\newblock \bibinfo{title}{Minimal reporting guideline for research involving eye tracking (2023 edition)}.
\newblock \emph{\bibinfo{journal}{Behavior research methods}} \bibinfo{pages}{1--7} (\bibinfo{year}{2023}).

\bibitem{sluch2017enhanced}
\bibinfo{author}{Sluch, V.~M.} \emph{et~al.}
\newblock \bibinfo{title}{Enhanced stem cell differentiation and immunopurification of genome engineered human retinal ganglion cells}.
\newblock \emph{\bibinfo{journal}{Stem cells translational medicine}} \textbf{\bibinfo{volume}{6}}, \bibinfo{pages}{1972--1986} (\bibinfo{year}{2017}).

\bibitem{Xypakis:23}
\bibinfo{author}{Xypakis, E.}, \bibinfo{author}{de~Turris, V.}, \bibinfo{author}{Gala, F.}, \bibinfo{author}{Ruocco, G.} \& \bibinfo{author}{Leonetti, M.}
\newblock \bibinfo{title}{Physics-informed deep neural network for image denoising}.
\newblock \emph{\bibinfo{journal}{Opt. Express}} \textbf{\bibinfo{volume}{31}}, \bibinfo{pages}{43838--43849} (\bibinfo{year}{2023}).
\newblock \urlprefix\url{https://opg.optica.org/oe/abstract.cfm?URI=oe-31-26-43838}.

\end{thebibliography}

\end{document}